\documentclass[]{spie}  

\usepackage{amsmath,amsfonts,amssymb}
\usepackage{graphicx}
\usepackage[colorlinks=true, allcolors=blue]{hyperref}

\title{Processing diffraction-limited images through innovative Super-Resolution techniques}

\author[a,d]{E. Quevedo}
\author[b,c]{S. Velasco}
\author[b,c]{C. Colodro-Conde}
\author[c]{G. Rodr\'iguez-Coira}
\author[b,c]{A. Oscoz}
\author[b,c]{R.L. L\'opez}
\author[b,c]{J. Font}
\author[a]{J. H. Brito}
\author[a]{O. Llin\'as}
\author[d]{S. Robaina}
\author[d]{G. M. Callic\'o}
\author[d]{R. Sarmiento}
\author[b,c]{R. Rebolo}

\affil[a]{Plataforma Oce\'anica de Canarias, Crtra. de Taliarte s/n., Las Palmas, 35214, Spain.}
\affil[b]{Instituto de Astrof\'isica de Canarias, c/ V\'ia L\'actea s/n, La Laguna, Tenerife E-38205, Spain.}
\affil[c]{Departamento de Astrof\'isica, Universidad de La Laguna, La Laguna, Spain.}
\affil[d]{Instituto Universitario de Microelectr\'onica Aplicada, Universidad de Las Palmas de Gran Canaria, Campus Universitario de Tafira s/n, Las Palmas, 35017, Spain.}

\authorinfo{Further author information: \\E.Q: E-mail: eduardo.quevedo@plocan.eu, equevedo@iuma.ulpgc.es}

\begin{document} 
\maketitle

\begin{abstract}
In the ELTs era, where the need for versatile and innovative solutions to produce very high spatial resolution images has become a major issue, the search of synergies with other science fields seems a logic step. 
One of the considered alternatives to reach high-resolution images is the use of several frames of the same target, this approach is known as fusion Super-Resolution in the state of the art.
Here, we propose the use of the super-resolution techniques based on structural similarity and initially developed for submarine environments\cite{Quevedo17}. Accordingly, innovative algorithms are implemented in order to process the science images from an Adaptive Optics system to obtain diffraction-limited images in the optical wavelengths.
\end{abstract}


\section{INTRODUCTION}
\label{sec:intro}  

The Strehl ratio produced by an AO instrument is not always the best possible. For long exposure times, some instruments such as VisAO \cite{Males10} implement a physical shutter to select only the best performance moments, acting as a frame selector. We are presenting here the proposed frame selection method for the state-of-the-art Adaptive Optics Lucky Imager (AOLI) \cite{Velasco16}\cite{Colodro17}. This instrument is designed to perform AO assisted LI observations to reach the diffraction limit on large size telescopes. However, the common LI algorithm based on the brightest pixel of the speckles pattern may not obtain the best possible result.

\newpage

The considered algorithm used as a starting point in this work improves the spatial resolution using Super-Resolution (SR) applied to Multi-Camera (MC) systems and dynamic imaging and has been developed by the Diseño de Sistemas Integrados (DSI) division at the Instituto Universitario de Microelectrónica Aplicada (IUMA) from Universidad de Las Palmas de Gran Canaria (ULPGC), in collaboration with PLataforma Oceánica de CANarias, (PLOCAN).

The SR algorithm consists on:

- Preprocessing: in this step the information is selected.

- Multicamera SR: Based on the overlapping between cameras.

- Selection filter: To select some specific frames and macro-blocks from the sample.

The SR algorithm is run on every frame in the sequence and can be done in real time even under a fast frame rate.

\section{SSIM with Airy Disk}

The basis of the proposed method consists on comparing the science target frames from the Adaptive Optics instrument (see Figure 1) with the theoretical diffraction-limited Airy Disk for the instrument's optical setup, taking the SSIM into consideration for this. By implementing SSIM selection into the Adaptive Optics processed observations procedure by the comparison with the Airy Disk, we have upgraded our method to obtain near diffraction-limited images even under bad seeing conditions. The probability of getting a better image considering the closure of the AO loop is shown in Figure 2.
\newline

\begin{figure}[h]
\centering
\includegraphics[width=\textwidth]{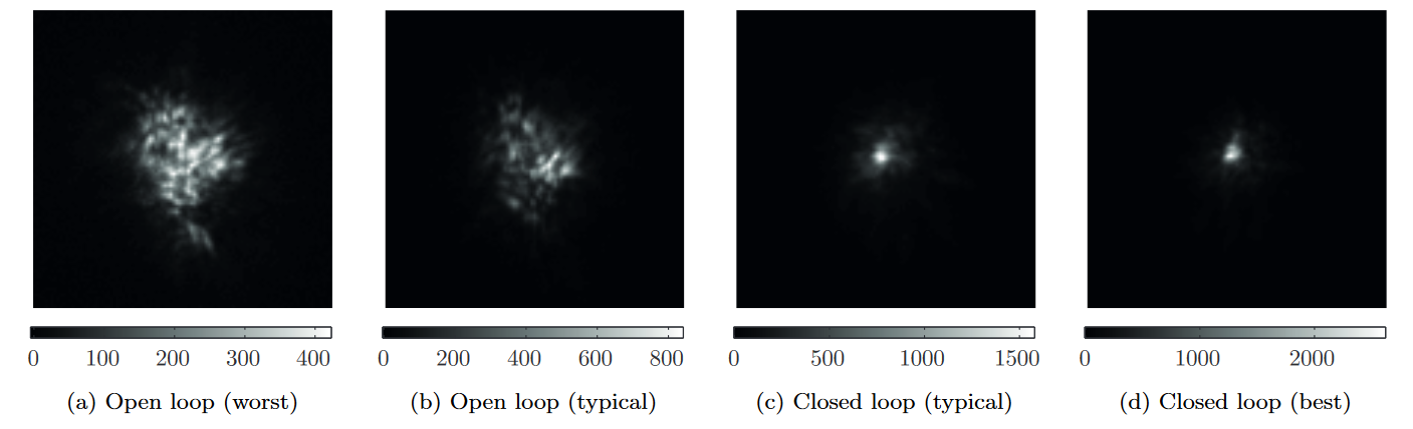}
\caption{Images of HIP10644 with and without AO corrections, showing worst, typical and best images according to the maximum pixel value.}
\end{figure}

\begin{figure}[h]
\centering
\includegraphics[width=.6\textwidth]{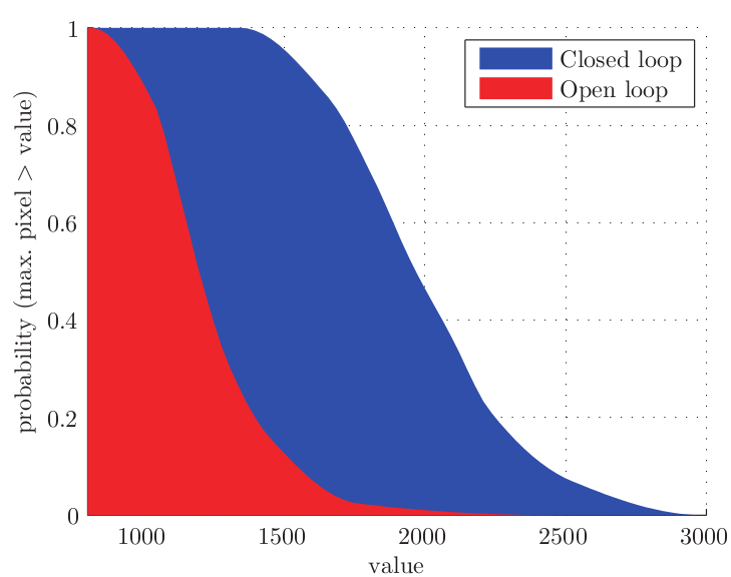}
\caption{Probability of obtaining an image whose maximum pixel value is higher than a given value (measure of quality). Closing the AO loop gives a higher probability of getting a better image.}
\end{figure}
\newpage 

The great advantage of this method comes from its post-processing nature. As all the frames are collected, even the not so good ones, we can generate an ordered list of AO corrected frames according to their similarity to the Airy disk. Figure 3 shows the enhancement of the Structural Similarity Index versus the classical brightest pixel method.

\begin{figure}[h]
\centering
\includegraphics[width=.9\textwidth]{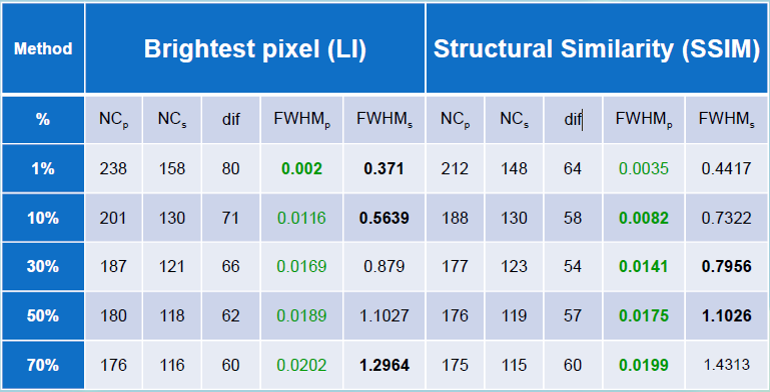}
\caption{LI and SSIM methods comparison.}
\end{figure}
\newpage

\section{Conclusions}

Applying the proposed SR method to some stellar sequences considering the aimilarity with the Airy Disk has increased the resolution of the final images appying Lucky Imaging. An additional point in favour of the SR method is that it has been demonstrated that it keeps the photometric information of the sources.



\bibliography{biblio} 
\bibliographystyle{spiebib} 

\end{document}